\documentclass[12pt,twoside]{article}
\usepackage{graphicx,amsmath,amssymb,latexsym}

\newcommand{\SM}{$SU(3)\times SU(2)_L\times U(1)_Y$ }

\begin{document}
\pagestyle{myheadings} \markboth{Grand Unification in Higher Dimensions with split-SUSY}{Grand Unification in
Higher Dimensions with split-SUSY}
\title{{\bf Grand Unification in Higher Dimensions with Split Supersymmetry}}
\author{Philip C. Schuster\footnote{Email: schuster@fas.harvard.edu}\\
                     \\
{\it Department of Physics} \\
{\it Harvard University} }
\date{\small\today}
\maketitle
\begin{abstract}
\noindent We investigate gauge coupling unification in higher dimensional GUT models with split supersymmetry. We
focus on 5d and 6d orbifold GUTs, which permit a simple solution to several problems of 4D GUTs as well as control
over GUT scale threshold corrections. In orbifold GUTs, calculable threshold corrections can raise or lower the
prediction for $\alpha_s(M_Z)$ in a way that depends on the location of Higgs fields. On the other hand, split
supersymmetry lowers the prediction for $\alpha_s(M_Z)$. Consequently, split supersymmetry changes the preferred
location of the Higgs fields in orbifold GUTs. In the simplest models, we find that gauge coupling unification
favors higgs doublets that live on the orbifold fixed points instead of in the bulk. In addition, relatively high
scales of supersymmetry breaking of $10^{10\pm 2}$ GeV are generically favored.
\end{abstract}
\section{Introduction}
Weak scale supersymmetry provides a nice solution to the naturalness problem and a predictive framework for electroweak symmetry breaking. It
also successfully predicts gauge couplings unification at a scale of $\approx 2\times 10^{16}$ GeV thereby providing support for the idea that
the standard model is embedded in a grand unified theory (GUT) \cite{4DGUT:Basics}. Recently, Arkani-Hamed and Dimopoulos proposed that
naturalness may not be a good criterion for determining weak scale physics \cite{Arkani-Hamed:2004fb}. Instead, they proposed that the higgs
mass is fine-tuned and looked for  alternative motivations for supersymmetry not tied to naturalness. Giudice and Romanino investigated this
possibility further and found that the MSSM with heavy scalar superpartners and light fermionic superpartners emerged naturally by demanding
gauge coupling unification and a viable dark matter candidate \cite{Giudice:2004tc}. This framework is known as split supersymmetry (split-SUSY)
and further phenomenological consequences and string theory realizations have been worked out in
\cite{Arkani-Hamed:2004yi,Antoniadis:2004dt,Sarkar:2004cs,Kokorelis:2004dc}.

In this paper, we focus on the issue of gauge coupling unification and the embedding of split-SUSY in a viable GUT model. We focus on orbifold
GUT models with a single extra dimension because these provide a particularly nice solution to many of the standard problems with 4d GUTs and a
calculable framework for high-scale threshold corrections
\cite{Kawamura:1999nj,Kawamura:2000ev,Kawamura:2000ir,Altarelli:2001qj,Kobakhidze:2001yk,Hall:2001pg}. In the simplest 4d supersymmetric SU(5)
GUTs, the prediction for the strong coupling constant at the scale $M_Z$ is $\alpha^{MSSM,GUT}_s(M_Z)=0.130\pm0.004$\footnote{We've neglected
including threshold contributions in quoting this prediction.} \cite{4DSusyPredictions}, somewhat larger than the experimentally measured value
of $\alpha_s^{exp}=0.119\pm0.002$ \cite{Eidelman:2004wy}. One of the important aspects of split-SUSY is a prediction for $\alpha_s(M_Z)$ that is
smaller than in 4d GUT scenarios. Likewise, one of the important features of orbifold GUTs is the presence of threshold corrections to gauge
couplings coming from heavy KK states that can improve the agreement between the predicted value of $\alpha_s(M_Z)$ and experiment. The primary
goal of this paper is to show how the competing effects on $\alpha_s(M_Z)$ from split-SUSY and 5d orbifold GUT thresholds constrain the
structure of generic orbifold GUTs. Contrary to the findings of previous authors \cite{Hall:2001pg,Hall:2001xb,Hall:2002ci,Hall:2002ea} that
bulk higgs fields are preferred for gauge coupling unification in low-energy SUSY models, we find that brane localized higgs fields are
naturally preferred in the split-SUSY scenario.

This result is attractive for several reasons. First, if the Higgs fields live on a brane, then the simplest possibility of just having the MSSM
or SM Higgs doublets can be realized \cite{Hebecker:2001wq}. Moreover, brane higgs scenarios are readily compatible with inherently 4d
mechanisms of electro-weak symmetry breaking. Of course, we do lose a few nice features with brane Higgs doublets such as unified quark-lepton
mass relations and charge quantization, but in any case SU(5) quark-lepton mass relations for the first two generations are seemingly
inconsistent with experiment.

Our paper is outlined as follows. In section 2, we'll review the framework of orbifold GUTs with extra dimensions.
Although the findings of our analysis will apply quite generally, we'll focus on the class of models developed in
\cite{Hall:2001pg,Hall:2001xb,Hall:2002ci,Hall:2002ea} to illustrate the important features of orbifold GUTs with
split-SUSY. In section 3, we'll discuss the effects of raising the SUSY breaking scale above $M_Z$ as well as the
competing effects of heavy KK state thresholds. In section 4, we'll present the results of our two-loop analysis
of gauge coupling unification and our findings for the favored range of the SUSY breaking scale and the location
of the higgs fields in an orbifold GUT completion. We then end with concluding remarks.
\section{GUTs with Extra Dimensions}
Among the many successes of 4d SUSY GUTs is the explanation for charge quantization and the pattern of quark and
lepton quantum number, a prediction for gauge coupling unification close to experimental bounds, quark-lepton mass
relations that reduce the number of flavor parameters in the standard model, and a robust framework for generating
small nonzero Majorana neutrino masses. However, important issues remain unresolved in these models. Chief among
these is the predicted rate of proton decay from colored higgsino exchange that essentially excludes the simplest
SU(5) GUTs. Other problems include the origin of SU(5) breaking, the very large splitting in mass required between
the Higgs colored triplet states and weak doublets, the fact that the observed quark/lepton mass relations for the
lighter generations violate generic GUT relations, and little explanation for the other flavor hierarchies of the
standard model.

GUT models with extra dimensions offer several possible resolutions to these problems. As in
\cite{Hall:2001pg,Hall:2001xb,Hall:2002ci,Hall:2002ea}, one can construct a 5d or higher dimensional model with
SU(5) gauge symmetry and break the symmetry down to \SM on 4d branes using boundary conditions. For example, in 5d
orbifold theories, orbifold fixed points serve as the branes on which the SM fields can live. The boundary
conditions emerge as a consequence of requiring the fields to transform with definite parity under the orbifold
group. With such a setup, the quark and lepton families and higgs of the standard model can be added to either the
bulk or to the branes.

The low energy effective theory consists of the lightest states in the KK expansion of fields. What should we
require of the low energy theory? In our case, we want to recover the MSSM at low energy. All of the SU(5) triplet
states should be heavy, and proton decay should be suppressed to experimentally acceptable levels. To see how
these requirements can be naturally satisfied in a higher dimensional GUT, let's consider the specific model
developed in \cite{Hall:2001xb}. Not only does this model illustrate the important features of higher dimensional
GUTs, but we will use this model to concretely study gauge coupling unification with split-SUSY later.

\subsection{A 5d Orbifold GUT Model}
Based on \cite{Hall:2001xb}, the model we'll consider contains a single extra dimension $S^1$ orbifolded under the
discrete reflection $Z:y\rightarrow -y$ and combined translation and reflection $Z^{\prime}:y\rightarrow -y+2\pi
R$. $\frac{S^1}{Z\times Z^{\prime}}$ has two orbifold fixed points (orbifold branes) located at $y=0$ and $y=\pi
R$ where $R$ is the radius of $S^1$ \cite{Hall:2001xb}. It is assumed that the bulk theory has 5d N=1 SUSY so that
there is a natural way to obtain 4d N=1 SUSY on the orbifold branes. SU(5) gauge fields are taken to reside in the
bulk so that boundary conditions can break the symmetry down to \SM on at least one of the two orbifold branes.
Thus, bulk gauge fields reside in a 5d N=1 SUSY vector multiplet, $(V,\Sigma)$, consisting of a 4d N=1 SUSY vector
multiplet $V$ and chiral multiplet $\Sigma$ transforming in the adjoint of SU(5). Matter fields can reside either
in the bulk or on the branes. Bulk matter fields reside in 5d N=1 SUSY hypermultiplets, $(\Phi,\Phi^c)$,
consisting of 4d N=1 SUSY chiral and anti-chiral multiplets $\Phi$ and $\Phi^c$ respectively.

Under the orbifold actions, the constituent multiplets transform with definite parity. To preserve a single N=1
SUSY in the zero mode spectrum, the gauge fields $V$ and $\Sigma$ are given $Z$ parities of $+$ and $-$
respectively. Bulk hypermultiplet constituents $\Phi$ and $\Phi^c$ are given parities $+$ and $-$ respectively.
$Z^{\prime}$ acts on the fundamental of SU(5) as $P_{Z^{\prime}}=(+,+,+,-,-)$. In addition, there can be extra
factors of $\eta_{\Phi}=\pm 1$ for bulk hypermultiplets. The above parity assignments lead to boundary conditions
on the fields at the orbifold fixed points,
\begin{eqnarray}\label{BoundaryConditions}
\nonumber V^{\pm}(x^{\mu},y) &=& V^{\pm}(x^{\mu},-y)=\pm V^{\pm}(x^{\mu},-y+2\pi R) , \\
\nonumber \Sigma^{\pm}(x^{\mu},y) &=& -\Sigma^{\pm}(x^{\mu},-y)=\pm \Sigma^{\pm}(x^{\mu},-y+2\pi R) ,  \\
\nonumber \Phi^{\pm}(x^{\mu},y) &=& \Phi^{\pm}(x^{\mu},-y)=\pm \eta_{\Phi}\Phi^{\pm}(x^{\mu},-y+2\pi R) ,  \\
 \Phi^{c\pm}(x^{\mu},y) &=& -\Phi^{c\pm}(x^{\mu},-y)=\pm \eta_{\Phi}\Phi^{c\pm}(x^{\mu},-y+2\pi R) ,
\end{eqnarray}
where the superscript $\pm$ refers to the parity under the SU(5) breaking action of $Z^{\prime}$. At $y=0$, we see
that bulk 5d N=1 (4d N=2) SUSY has been broken down to 4d N=1 by $Z$, but SU(5) is still operative. At $y=\pi R$,
the bulk SUSY has been broken down to 4d N=1 and SU(5) has been broken with only \SM surviving.

A single generation of standard model fermions can live in the bulk if they reside in two hypermultiplets
transforming in the {\bf 10}, $T(u,e)$ and $T^{\prime}(q)$, and two in the ${\bf \bar{5}}$, $F(d)$ and
$F^{\prime}(l)$. With the choice $\eta_T=\eta_F=1$ and $\eta_{T{\prime}}=\eta_{F^{\prime}}=-1$, the zero modes of
these four hypermultiplets fill out a single generation of standard model fermions. There is also the possibility
of part of a generation living on an orbifold brane with the remainder in the bulk. For example, $d$ and $l$ can
come from a 4d N=1 susy ${\bf \bar{5}}$ multiplet living on a brane while the $u$, $e$, and $q$ components come
from two {\bf 10}s in the bulk.

If the Higgs fields arise from the bulk, they can come from two bulk hypermultiplets transforming in the {\bf 5}
and ${\bf \bar{5}}$ \cite{Hall:2001xb}. Another possibility is for the Higgs to arise from a vector multiplet. For
example, Higgs fields with the correct quantum numbers can come from the doublet components of the adjoint of
SU(6) under its \SM decomposition. 6d orbifold models with this feature have been constructed in
\cite{Hall:2001zb}. Of course, the Higgs can also be a brane field \cite{Hebecker:2001wq}.

Having introduced the orbifold GUT framework, we can see how the problems of standard 4d SUSY GUTs can be
resolved. The Higgs doublet-triplet splitting problem is solved by requiring boundary conditions that eliminate
the triplet zero mode. The Higgs triplet states are now naturally heavy with mass of order, $1/M_c$, where
$M_c=1/R$ is the scale of the extra dimension. We also have the possibility of placing Higgs fields on the 4d
SU(5) violating brane in which case there are not necessarily any triplet partners to begin with.

Yukawa couplings of bulk hypermultiplets are forbidden by 5d supersymmetry and so the yukawa couplings reside on
the branes. Thus, if the Higgs is in the bulk, then proton decay from dimension five triplet higgsino exchange via
dirac mass terms is eliminated by the bulk supersymmetry. Additional dangerous sources of proton decay can be
eliminated by using the bulk SU(2)$_R$ symmetry that comes from the N=2 4d SUSY of the 5d bulk \cite{Hall:2001xb}.

SU(5) mass relations can be preserved for the heaviest generation by placing it on the SU(5) brane at $y=0$. If
the first two generations are placed in the bulk, then their masses will not respect SU(5) relations because the
down-type quarks and charged leptons have different yukawa couplings to the Higgs field for our choice of
representation. Moreover, because the bulk fields are spread out in the extra dimension, wave-function suppression
will naturally make their masses smaller thereby explaining why heavy matter fields satisfy SU(5) mass relations
while light matter does not.

Gauge coupling unification can proceed as usual except now there will be radiative corrections coming from KK
modes and brane localized gauge kinetic operators that do not respect the bulk SU(5). As long as the extra
dimension is large compared to the unification scale, then the bulk gauge kinetic operators will dominate over
brane localized operators by a factor of $\frac{M_s}{M_c}$. Of course, this assumes that we can reliably estimate
the couplings to be of comparable strength at some scale. We will later identify the unification scale with the
scale of strong coupling for the 5d theory thereby justifying this assumption.

In this paper, we will also briefly consider 6d SO(10) orbifold GUTs. SO(10) orbifold GUTs on $\frac{T^2}{Z_2}$
have been constructed and their features discussed in the literature \cite{Hall:2001xr}. Other 6d orbifold GUTs
with unified gauge group SU(6) have been constructed that contain Higgs doublets arising from bulk gauge fields
\cite{Hall:2001zb}. For our purposes, the primary impact on gauge coupling unification that 6d models introduce
consists of different power law scaling of couplings above the scale $M_c$ than in 5d models. As will be made
explicit in section 3, the primary effect of a sixth dimension will be to decrease the overall magnitude of the
contribution to $\alpha_s$ from KK thresholds.

\section{Gauge Coupling Unification in Split-SUSY}
In this section, we discuss gauge coupling unification in split-SUSY. We start by considering the experimentally measured values,
$\sin^2\theta_W(M_Z)=0.23150\pm 0.00016$, $\alpha^{-1}(M_Z)=128.936\pm 0.0049$, and $\alpha_s(M_Z)=0.119\pm 0.003$ \cite{Eidelman:2004wy}. Given
$\sin^2\theta_W(M_Z)$ and $\alpha^{-1}(M_Z)$, we can obtain a prediction for $\alpha_s(M_Z)$ assuming unification at a high scale. A one-loop
analysis of this prediction will give a prediction for $\alpha_s(M_Z)$ with errors dominated by the large SU(3) coupling of order
$(\alpha_s^{\mbox{1-loop}}(M_Z))^2$. However, the experimental uncertainty is of order $.003\approx (\alpha_s(M_Z))^3$, so a full two-loop
analysis with one-loop thresholds is needed to reliably compare theory with experiment.

The easiest way to calculate gauge coupling predictions is to use a succession of effective field theories (EFTs)
obtained by integrating out heavy particles at the appropriate mass scales \cite{Hall:1980kf}. In this way, we can
use a simple mass independent renormalization scheme such as $\bar{MS}$ (or $\bar{DR}$ if we're working with SUSY)
in each effective theory. The effects of large log contributions at lower energies coming from massive states is
absorbed into the matching conditions between the theories. Assuming that the gauge couplings are unified at a
scale $M_G$, the evolution of the coupling down to $M_Z$ proceeds in two steps. In the first step, the tower of KK
modes contribute at one-loop to the gauge couplings above the compactification scale $M_c$. From the 5d
perspective, the theory is not renormalizable which is reflected by the mass dimension of the 5d gauge coupling.
Thus, we expect power-law scaling of the gauge couplings between the GUT scale $M_G$ and $M_c$. In the second
step, the gauge couplings run in the usual logarithmic fashion from $M_c$ to $M_Z$.

For clarity, we start by discussing the familiar logarithmic running and matching below the scale $M_c$. After
studying the one-loop effects of lifting the SUSY breaking scale $m_S$, we will discuss the scaling of the
couplings above $M_c$.

\subsection{Running and Matching Gauge Couplings}
The underlying UV theory is the model presented in section 2. After matching between the 5d theory and the
effective 4d theory at the compactification scale $M_c$, we obtain the MSSM. In the split-SUSY scenario, we assume
that the squarks, sleptons, charged and pseudoscalar Higgs are degenerate with mass $m_S$. Below the scale $m_S$,
the effective theory consists of only the higgsinos $\tilde{H}_{u,d}$, gluinos $\tilde{g}^{\alpha}$, W-inos
$\tilde{W}^a$, B-ino $\tilde{B}$, and the standard model fields with a single higgs doublet $H$,
\begin{eqnarray}\label{SplitSusyLagrangian}
\nonumber \mathcal{L}_{SSSM} &=& \mathcal{L}_{gauge} + m^2H^{\dagger}H-\frac{\lambda}{2}(H^{\dagger}H)^2 \\
\nonumber &-& [y^u_{ij}\bar{q}_ju_i(i\sigma_2H^*) + y^d_{ij}\bar{q}_jd_iH+y^e_{ij}\bar{l}_je_iH \\
\nonumber &+& \frac{M_3}{2}\tilde{g}^{\alpha}\tilde{g}^{\alpha} + \frac{M_2}{2}\tilde{W}^a\tilde{W}^a
+ \frac{M_1}{2}\tilde{B}\tilde{B} \\
\nonumber &+& \mu\tilde{H}_u^T(i\sigma_2\tilde{H}_d)
          + \frac{1}{\sqrt{2}}H^{\dagger}(\tilde{g}_u\sigma^aW^a+\tilde{g}^{\prime}_u\tilde{B})\tilde{H}_u \\
          &+&\frac{1}{\sqrt{2}}(H^Ti\sigma_2)(-\tilde{g}_d\sigma^a\tilde{W}^a+\tilde{g}^{\prime}_d\tilde{B})H_d] + h.c.
\end{eqnarray}
With $H_u$ and $H_d$ the up-type and down-type MSSM Higgs doublets respectively, we fine tune the linear
combination $H=-\cos(\beta)i\sigma_2H^*_d+\sin(\beta)H_u$ to be light. The tree level matching condition between
the MSSM and the split-susy lagrangian can be found by taking $H_u\rightarrow \sin(\beta)H$ and
$H_d\rightarrow\cos(\beta)i\sigma_2H^*$ in the MSSM lagrangian, so that
\begin{eqnarray}\label{SplitSusyMatching}
\nonumber \lambda(m_S) &=& \frac{g_2^2+\frac{3}{5}g_1^2}{4}\cos^2(2\beta),\\
\nonumber y^u_{ij} &=& \lambda^u_{ij}\sin(\beta),\\
\nonumber y^{d,e}_{ij} &=& \lambda^{d,e}_{ij}\cos(\beta),\\
\nonumber \tilde{g}_u &=& g_2\sin(\beta),\\
\nonumber \tilde{g}_d &=& g_2\cos(\beta),\\
\nonumber \tilde{g}^{\prime}_u &=& \sqrt{\frac{3}{5}}g_1\sin(\beta),\\
  \tilde{g}^{\prime}_d &=& \sqrt{\frac{3}{5}}g_1\cos(\beta),
\end{eqnarray}
where $\lambda^{u,d,e}_{ij}$ are the Higgs yukawa coupling matrices in the MSSM. To obtain predictions for
standard model couplings at the weak scale $M_Z$, we also need to match the split-SUSY theory onto a low energy
theory containing just the standard model fields \cite{Arason:1991ic}.


\begin{figure}
\begin{center}
\setlength{\unitlength}{1cm}
\begin{picture}(8,7)
\thicklines \put(1,0){\line(1,0){3}}\put(4.5,0){\large{TeV}}\put(1,1.5){\large{Split-SUSY}}
\put(1,3){\line(1,0){3}}\put(4.5,3){\large{$m_S$}}\put(1,3.5){\large{MSSM}}
\put(1,4.5){\line(1,0){3}}\put(4.5,4.5){\large{$M_c^{\prime}$}}\put(1,5){\large{5d SU(5) GUT}}
\put(1,6){\line(1,0){3}}\put(4.5,6){\large{$M_G$}}\put(1,6.5){\large{UV Completion}}
\put(0,2){\vector(0,3){3}}\put(-1,1.7){Energy}
\end{picture}
\end{center}
\caption{Summary of the effective theories and their relevant scales in a 5d SU(5) orbifold GUT with
split-SUSY}\label{fig:EnergyTower}
\end{figure}
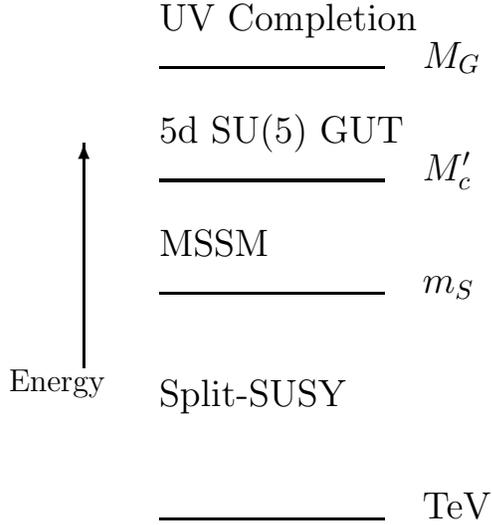

With our collection of effective theories (illustrated in figure \ref{fig:EnergyTower}), we RG run the MSSM from
the compactification scale $M_c$ to the scale of supersymmetry breaking $m_S$ \cite{Giudice:2004tc}. We then match
onto the split-SUSY effective lagrangian (\ref{SplitSusyLagrangian}) using the matching conditions
(\ref{SplitSusyMatching}). After RG running the split-susy couplings to the scale of the higgsino and gaugino
masses, we match onto the standard model. Finally, we match the low-energy parameters to experimentally measured
couplings and extract a prediction for $\alpha_s(M_Z)$ using measured values of $\sin^2\theta_W(M_Z)$,
$\alpha(M_Z)$, and masses of the standard model fermions as inputs.

In matching the theories above and below $m_S$, we must consider several large threshold corrections. First, the
running of the gaugino and higgsino masses can lead to a significant spread in masses at low energy, so separate
thresholds are included for each. The heavy top mass also contributes a large threshold above the weak scale, so
it is also included.\footnote{The difference between the $\bar{MS}$ running and pole mass can lead to additional
threshold correction that should also be included. In our analysis, this difference is significant only for the
gluinos \cite{Giudice:2004tc}.} We should in principle include all of the one-loop thresholds coming from the
sparticle and higgs scalar spectrum at $m_S$, but it is a decent approximation to ignore these effects given that
our theory is weakly coupled at $m_S$ for $m_S$ sufficiently large.

\subsection{One-Loop Analysis}
Using the RGEs in \cite{Giudice:2004tc}, we can obtain some insight into the consequences of split-SUSY and GUT
thresholds for gauge coupling unification by calculating their one-loop effects. First, we focus on the effects of
lifting $m_S$ to a high scale and so we momentarily ignore GUT thresholds. The gauge couplings $\alpha_i(M_Z)$ can
be written as
\begin{eqnarray}\label{OneLoopRunning}
\nonumber \frac{1}{\alpha_i(M_Z)} &=& \frac{1}{\alpha_i(M_G)}+\frac{b^{MSSM}_i}{2\pi}\log{\frac{M_G}{M_Z}} \\
                                  &+&
\frac{b_i^{SSSM}-b_i^{MSSM}}{2\pi}\log{\frac{m_S}{M_Z}}+\frac{b_i^{SM}-b_i^{SSSM}}{2\pi}\log{\frac{M}{M_Z}}
                                  + \gamma_i + \delta_i ,
\end{eqnarray}
where $M_G$ is some high mass scale (either the compactification scale, or if we were doing conventional
unification the GUT scale), $M$ is the mass scale of gauginos and higgsinos, and
$b_i^{SM}=(\frac{41}{10},-\frac{19}{6},-7)$, $b_i^{MSSM}=(\frac{33}{5},1,-3)$, and
$b_i^{SSSM}=(\frac{9}{2},-\frac{7}{6},-5)$ are the $\beta$-function coefficients for the standard model, the MSSM,
and split-SUSY respectively. The two-loop contributions are contained in the $\gamma_i$ factors and additional
small threshold factors are included in $\delta_i$. In the limit $m_S=M=M_Z$, an easy way to obtain a prediction
for $\alpha_s(M_Z)=\alpha_3(M_Z)$ is to take the linear combination
$\alpha^{-1}_3-\frac{12}{7}\alpha^{-1}_2+\frac{5}{7}\alpha^{-1}_1$. Using (\ref{OneLoopRunning}), the
$\log{\frac{M_G}{M_Z}}$ term cancels out and we are left with
\begin{eqnarray}\label{AlphaSResult}
\nonumber \alpha^{-1}_3(M_Z)&=&\frac{12}{7}\alpha^{-1}_2(M_Z)-\frac{5}{7}\alpha^{-1}_1(M_Z) \\
&+&\gamma_3-\frac{12}{7}\gamma_2+\frac{5}{7}\gamma_1+\delta_3-\frac{12}{7}\delta_2+\frac{5}{7}\delta_1,
\end{eqnarray}
where we've assumed that $\alpha_i(M_G)\approx \alpha(M_G)$. Feeding in the experimentally measured values of
$\alpha_{1,2}(M_Z)$, one obtains $\alpha^{MSSM,GUT}_s(M_Z)=0.130\pm0.004$ for the MSSM \cite{4DSusyPredictions}.

When $m_S, M\geq M_Z$, we can take the same linear combination as in (\ref{AlphaSResult}) to calculate
$\alpha_s(M_Z)$. The difference $\delta\alpha_s$ between $\alpha^{MSSM,GUT}_s(M_Z)$ and the value calculated in
the split-SUSY case is then approximately,
\begin{equation}\label{AlphaDifferenceA}
    \delta\alpha_s(M_Z)\approx -\frac{\alpha_s(M_Z)^2}{2\pi}\Delta,
\end{equation}
where $\Delta$ is given by,
\begin{equation}\label{AlphaDifferenceB}
    \Delta=(\frac{3}{14})\log{\frac{m_S}{M_Z}}+(\frac{8}{7})\log{\frac{M}{M_Z}} .
\end{equation}
Already we can see that raising the scale $m_S$ or making gauginos and higgsinos heavier lowers the prediction for
$\alpha_s(M_Z)$. Thus, split-SUSY can improve the agreement between $\alpha_s^{exp}=0.119\pm0.002$
\cite{Eidelman:2004wy} and $\alpha^{MSSM,GUT}_s$ calculated assuming unification.

\subsection{KK Contributions and GUT Thresholds}
Near the compactification scale $M_c$, loops of KK modes lead to SU(5) universal power-law scaling of the
couplings. There is also logarithmic non-universal running due to 4d brane kinetic terms and an effective zero
mode mismatch that we will discuss shortly. As in section 2, we assume that $M_c$ is sufficiently smaller than
$M_G$ to suppress brane kinetic contributions. Above $M_c$, the gauge couplings quickly become strong, so it is
natural to assume that the unification scale coincides with strong coupling. If we have $d$ extra dimensions, this
assumption fixes the ratio $\frac{M_G}{M_c}$ as
\begin{equation}\label{StrongCouplingAssumption}
(\frac{M_G}{M_c})^d\approx \frac{16\pi^2}{Cg^2} ,
\end{equation}
where $g^2$ is evaluated at the compactification scale and $C$ is a group theory factor ($C=5$ for SU(5) or $C=8$
for SO(10)) \cite{Hall:2001xb}. The strong coupling assumption helps us justify the use of NDA to estimate the
unknown threshold contributions coming from $M_G$. NDA also implies that brane gauge kinetic operators are
suppressed relative to bulk gauge kinetic operators by $\frac{M_c}{M_G}$ for a single extra dimension. $M_G$ scale
threshold corrections to the $\alpha_i^{-1}$ should naturally be of order $\approx \frac{1}{4\pi}$. Thus, we
expect the precision of our prediction for $\alpha_s(M_Z)$ to be limited by $\approx
\frac{\alpha_s(M_Z)^2}{4\pi}\sqrt{3}\approx 0.002$. In addition, the effects of strong coupling over a small
energy interval near $M_G$ can be expected to make contributions of threshold size. In all, we estimate the
uncertainty from GUT thresholds in our final calculation of $\alpha_s$ to be $\approx \pm 0.003$.

Assuming unification at $M_G$, the matching condition between the full theory and the 4d theory below $M_c$ is,
\begin{eqnarray}\label{McMatching}
\frac{1}{\alpha_i(M_c^{\prime})}=\frac{1}{\alpha_G}+\frac{c}{2\pi}[\frac{M_G}{M_c^{\prime}}-1]+
                                 \frac{\tilde{b}_i}{2\pi}\log{\frac{M_G}{M_c^{\prime}}}
                                 + \frac{\Delta^{KK,thr}_i}{2\pi},
\end{eqnarray}
where $M_c^{\prime}$ is the appropriate matching scale ($M_c^{\prime}=\frac{M_c}{\pi}$ in the model of section 2),
$c$ reflects the contributions from the tower of KK modes that lead to universal power-law scaling, $\tilde{b}_i$
are non-universal $\beta$-function coefficients, and $\Delta^{KK,thr}_i$ are threshold contributions. The
thresholds $\Delta^{KK,thr}_i$ come from integrating out the gauge and matter KK modes and can be calculated given
a choice for the matter representation and bulk geography. The bulk SU(5)-universal scaling controlled by $c$ can
be calculated given a suitable UV completion, but we will not bother with this because unification is not altered
by this scaling. As in \cite{Hall:2001pg,Hall:2001xb}, we choose $M_c\approx 10^{15}$ GeV so that the couplings
$\alpha_i$ are very nearly unified at $M_c$ before scaling to their unified value of $\approx 4\pi$.

In our analysis, the contributions from non-universal running are absorbed into a matching condition between
$\alpha_3$, $\alpha_2$, and $\alpha_1$ at the scale $M_c^{\prime}$. Again taking the linear combination
$\alpha^{-1}_3-\frac{12}{7}\alpha^{-1}_2+\frac{5}{7}\alpha^{-1}_1$, we obtain
\begin{equation}\label{McMatchingB}
\alpha_s^{-1}(M_c^{\prime})=\frac{12}{7}\alpha_2^{-1}(M_c^{\prime})-\frac{5}{7}\alpha_1^{-1}(M_c^{\prime})+\frac{\Delta_{KK}}{2\pi}
,
\end{equation}
where $\Delta_{KK}$ is
\begin{equation}\label{DeltaKK}
\Delta_{KK}=(\tilde{b}_3-\frac{12}{7}\tilde{b}_2+\frac{5}{7}\tilde{b}_1)\log{\frac{M_G}{M_c^{\prime}}} +
\Delta_{KK}^{thr},
\end{equation}
and $\Delta_{KK}^{thr}=(\Delta^{KK,thr}_3-\frac{12}{7}\Delta^{KK,thr}_2+\frac{5}{7}\Delta^{KK,thr}_1)$. From this
relation, we obtain an additional contribution to $\delta\alpha_s(M_Z)$ of,
\begin{equation}\label{AlphaDifferenceC}
\delta\alpha_s(M_Z)\approx -\frac{\alpha_s(M_Z)^2}{2\pi}\Delta_{KK} .
\end{equation}
So, for $\Delta_{KK}\geq 0$, the prediction for $\alpha_s(M_Z)$ is lowered, while for $\Delta_{KK}\leq 0$ it is
increased. This effect, and the competing effect in eq. (\ref{AlphaDifferenceA}) will determine the favored
location for the SM higgs fields in our toy model as well as the preferred range of SUSY breaking mass scales.

\begin{figure}
\begin{center}
\setlength{\unitlength}{1cm}
\begin{picture}(8,4)
\thicklines \put(3.5,0){\line(1,0){3}}\put(7,0){\large{$M_c^{\prime}$}}\put(-2.5,0.2){Calculable KK thresholds}
\put(3,.2){$\delta(\frac{1}{\alpha_i})=\frac{\Delta^{KK,thr}_i}{2\pi}\approx \frac{1}{2\pi}$}
\put(-2.5,1.8){Non-universal running}
\put(2.8,1.8){$\delta(\frac{1}{\alpha_i})=\frac{\tilde{b}_i}{2\pi}\log{\frac{M_G}{M_c^{\prime}}}\approx
\frac{3-5}{2\pi}$} \put(3.5,3){\line(1,0){3}}\put(7,3){\large{$M_G$}} \put(-2.5,3.2){Unknown thresholds}
\put(3,3.4){$\delta(\frac{1}{\alpha_i})\approx \frac{1}{4\pi}$} \put(8,0){\vector(0,3){3}}\put(8.4,0){Energy}
\end{picture}
\end{center}
\caption{Summary of threshold corrections and non-universal running between $M_c^{\prime}$ and
$M_G$.}\label{fig:KKCorrections}
\end{figure}
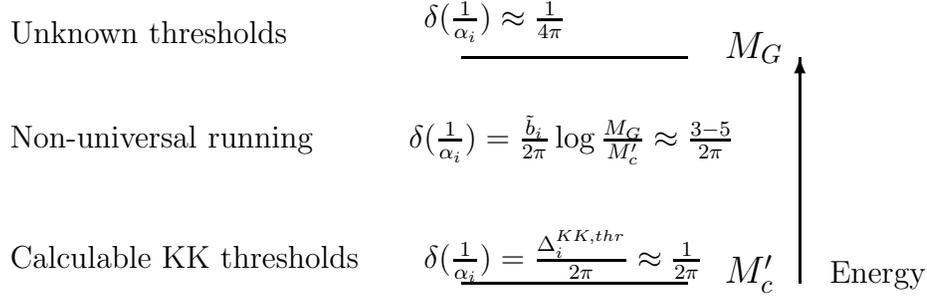

The simplest way to calculate the non-universal $\beta$-function coefficients $\tilde{b}_i$ is to think about what
the boundary conditions do to the spectrum of the bulk theory \cite{Hall:2001xb}. For example, consider the simple
case where the extra dimension is a circle $S^1$ orbifolded with a discrete $Z_2$ symmetry, $y\rightarrow -y$. The
boundary conditions on the bulk fields at the fixed point of $Z_2$ result from their $Z_2$ parities and divide the
KK tower into states that are even and odd under $Z_2$. The full KK tower of states on $S^1$ has completely SU(5)
invariant running because there are no SU(5) violating defects in the bulk. Under the orbifold map,
$S^1\rightarrow \frac{S^1}{Z_2}$, half of the states are projected out (i.e. left and right moving states are
mapped onto single states). For a particular bulk state $T$ with $\beta$-function coefficients $b_i$, let $T^O_n$
($n\geq 0$) and $T^E_n$ ($n\geq 0$) be the odd and even modes with masses $m^O_n=\frac{n+1/2}{R}$ and
$m^E_n=\frac{n}{R}$ respectively. The contribution from the odd states $T^O_n$ is therefore equivalent to a tower
of states $\tilde{T}^O_n$ ($\infty\geq n \geq -\infty$) on $S^1$ with beta function coefficients $b_i/2$. For the
even states, we can do the same except that the zero mode will not in general have a $\beta$-function coefficient
$b_i/2$. So, for the even modes, we can construct a tower of states equivalent to an $S^1$ tower with an effective
zero mode with coefficient $b^0_i-b_i/2$, where $b^0_i$ is the actual zero mode $\beta$-function coefficient. It
is precisely this effective zero mode that generates the one-loop non-universal running.

This analysis can be easily generalized. If a manifold $F$ is orbifolded under $M$ such that, apart from the zero
mode, $n_M$ states of $F$ are mapped to $\frac{F}{M}$, then the $\tilde{b}_i$ are given by,
\begin{equation}\label{MismatchEquation}
    \tilde{b}_i=b^0_i-\frac{b^{KK}_i}{n_M} ,
\end{equation}
where $b^0_i$ is the zero mode $\beta$-function coefficient and $b_i^{KK}$ is the the $\beta$-function coefficient
of the excited KK states in the tower connected to the zero mode. For the parity assignments of the model in
section 2, the KK tower connected to the zero modes has states with the same quantum numbers as the zero modes but
with N=1 5d SUSY (i.e. N=2 4d SUSY). For example, suppose the zero mode fills out a N=1 SU(m) vector field $V$
coming from a 5d N=1 vector $\{V,\Sigma\}$. The KK modes connected to the zero mode will then have a one-loop
$\beta$-function coefficient, $b^{KK}_{SU(m)}=-2m$. On the other hand, the zero mode $\beta$-function coefficient
is, $b^0_{SU(m)}=-3m$. So, $b^{KK}_i=\frac{2}{3}b^0_i$ and $\tilde{b}_i=\frac{2}{3}b^0_i$. Generally,
$\tilde{b}_i=\frac{2}{3}b^0_i$ for a N=1 SUSY vector $V$ zero mode that comes from a bulk 5d SUSY vector in the
adjoint. Analogous calculations show that, $b^{KK}_i=-2b_i^0$, and $b^{KK}_i=+2b_i^0$ when the zero mode is an
adjoint $\Sigma$, or a fundamental $\Phi$ respectively. Thus, with $n_M=n$, we have
$\tilde{b}_i=(1-\frac{2}{3n})b_i^0$, $\tilde{b}_i=(1+\frac{2}{n})b_i^0$, and $\tilde{b}_i=(1-\frac{2}{n})b_i^0$
when the zero mode is a $V$, $\Sigma$, or $\Phi$ respectively.

With the above result, we can explicitly calculate the $\tilde{b}_i$ in our case. First, the gauge fields come
from a bulk vector field, and the zero mode $\beta$-function coefficients are given by
$(b^0_1,b^0_2,b^0_3)=(0,-6,-9)$ for the MSSM. Thus,
$(\tilde{b}_1,\tilde{b}_2,\tilde{b}_3)=(0,-6+\frac{4}{n},-9+\frac{6}{n})$ for the gauge fields and so
$\tilde{b}_{gauge}=\frac{1}{7}(9-\frac{6}{n})$. The zero mode matter fields all come in complete SU(5) multiplets,
so they do not contribute to non-universal running even if they come from bulk fields. As for the Higgs fields,
they can come from brane fields, bulk hypermultiplet fields, or bulk gauge fields with the result that
$\tilde{b}_{higgs,brane}=-\frac{9}{7}$, $\tilde{b}_{higgs,hyper}=-\frac{9}{7}+\frac{18}{7n}$, or
$\tilde{b}_{higgs,gauge}=-\frac{9}{7}-\frac{18}{7n}$ respectively. In all, we therefore have,
\begin{eqnarray}\label{DeltaKKResults}
\nonumber  \Delta_{KK} &=& (\frac{-6}{7n})\log{\frac{M_G}{M_c^{\prime}}}  \quad (\mbox{brane localized Higgs}),\\
\nonumber  \Delta_{KK} &=& (\frac{12}{7n})\log{\frac{M_G}{M_c^{\prime}}}  \quad (\mbox{Higgs from bulk hypermultiplets}),\\
           \Delta_{KK} &=& (\frac{-24}{7n})\log{\frac{M_G}{M_c^{\prime}}} \quad (\mbox{Higgs from bulk vector multiplets}).
\end{eqnarray}
In the above, we've neglected $\Delta^{KK,thr}_i$ contributions, but we included them in our final two-loop
analysis. For reference, $\Delta_{KK}^{thr}=0.84$ for the brane and bulk hypermultiplet Higgs cases, and
$\Delta_{KK}^{thr}=-1.68$ for the bulk vector multiplet Higgs case \cite{Contino:2001si}.

Recalling that $\Delta_{KK}\geq 0$ lowers the prediction for $\alpha_s(M_Z)$, we see that Higgs fields coming from
bulk hypermultiplets lower $\alpha_s(M_Z)$ while brane and bulk vector multiplet Higgs fields increase
$\alpha_s(M_Z)$.

\section{Results for Gauge Coupling Unification}
In the spirit of the one-loop analysis above, we performed a two-loop analysis using the model of section 2 to
quantify what the unification predictions for $\alpha_s$ teach us about the bulk "geography" of the theory in
light of the split-SUSY scenario. In particular, we investigated the preferred scale of scalar superpartner masses
in split-SUSY and the preferred location of the Higgs fields in extra dimensional GUTs illustrated by this model.
In order to be as model independent as possible, we assumed complete mass degeneracy of the squark, slepton, and
charged and pseudoscalar Higgs. We also neglected one-loop thresholds at $m_S$. As is often done, we included only
the heaviest generation of SM quarks and leptons in the RG running of the gauge couplings above $M_Z$. We also
assumed higgsino and gaugino mass unification at $M_c^{\prime}$. Our results are given for higgsino and gaugino
unified mass boundary conditions of $300$ GeV and $1000$ GeV.

\begin{figure}
\includegraphics[width=12cm]{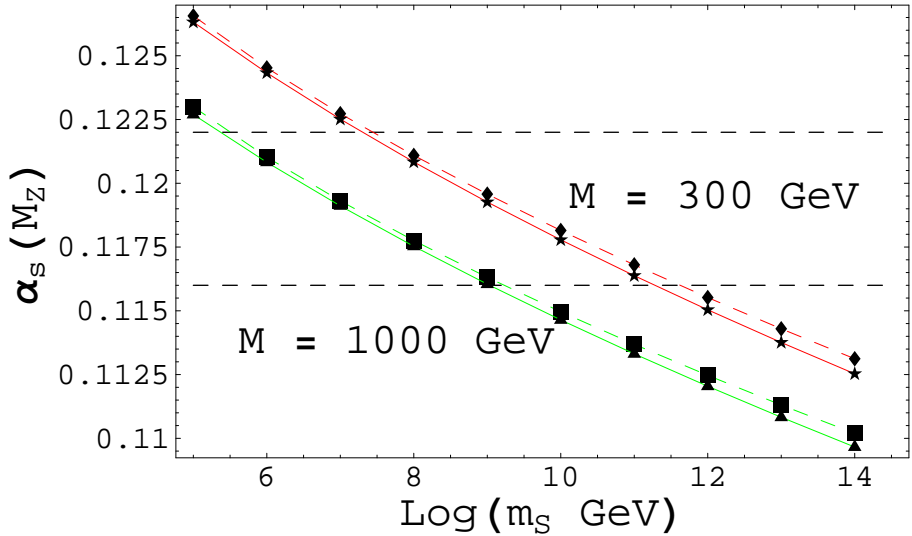}
\caption{The prediction for $\alpha_s(M_Z)$ as a function of the SUSY breaking scale $m_S$ for the 4D MSSM. The
horizontal dashed lines show the $1\sigma$ experimental constraint for
$\alpha_s^{exp}(M_Z)$\cite{Eidelman:2004wy}. The solid lines correspond to $\tan(\beta)=50$ and the dashed lines
to $\tan(\beta)=1.5$. We assume higgsino and gaugino mass unification at the unification scale.}
\label{fig:4DStrCou}
\end{figure}

\begin{figure}
\includegraphics[width=12cm]{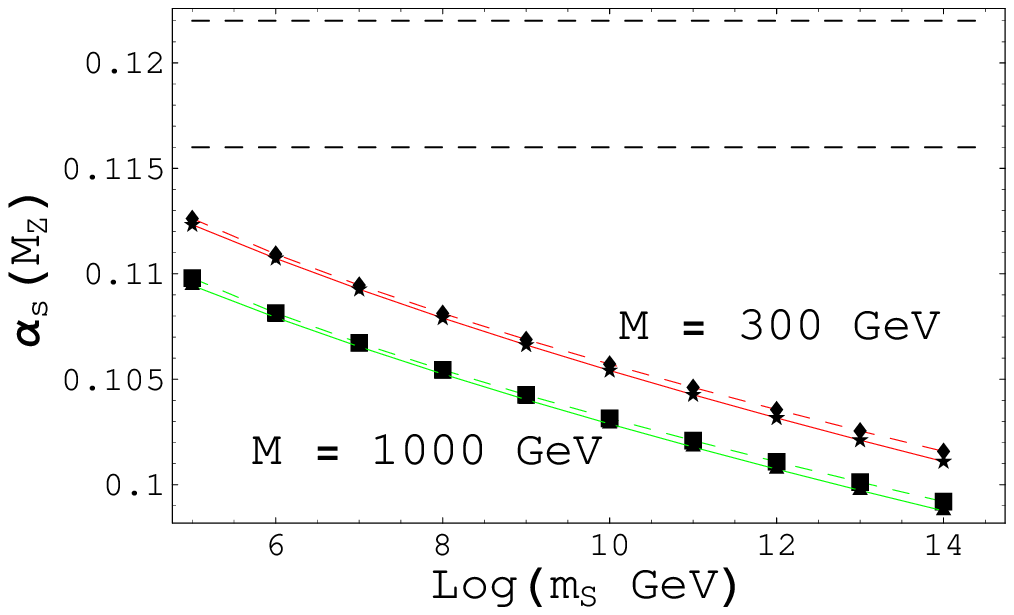}
\includegraphics[width=12cm]{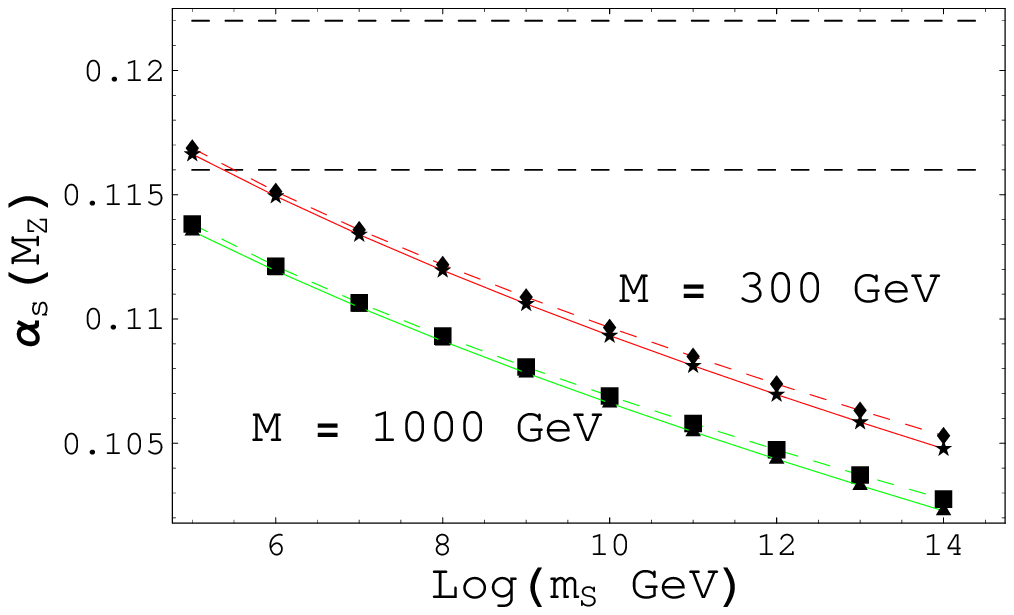}
\caption{The prediction for $\alpha_s(M_Z)$ as a function of the SUSY breaking scale $m_S$. The top graphic is for
the 5D SU(5) model with bulk hypermultiplet Higgs fields and the bottom graphic is for a 6D SO(10) model with bulk
hypermultiplet Higgs fields. The horizontal dashed lines show the $1\sigma$ experimental constraint for
$\alpha_s^{exp}(M_Z)$\cite{Eidelman:2004wy}. The solid lines correspond to $\tan(\beta)=50$ and the dashed lines
to $\tan(\beta)=1.5$. The compactification scale is set to $M_c^{\prime}=4\times 10^{14}$ GeV. We assume higgsino
and gaugino mass unification with mass $M$ at the compactification scale. Both $M=300$ GeV and $M=1000$ GeV are
shown.} \label{fig:5Dand6DHyperStrCou}
\end{figure}

\begin{figure}
\includegraphics[width=12cm]{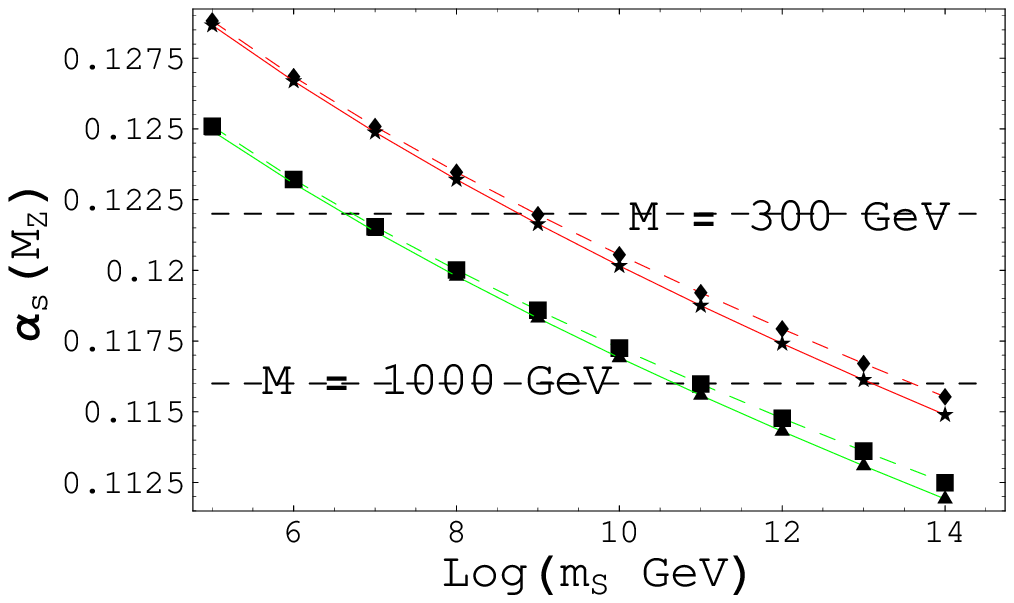}
\includegraphics[width=12cm]{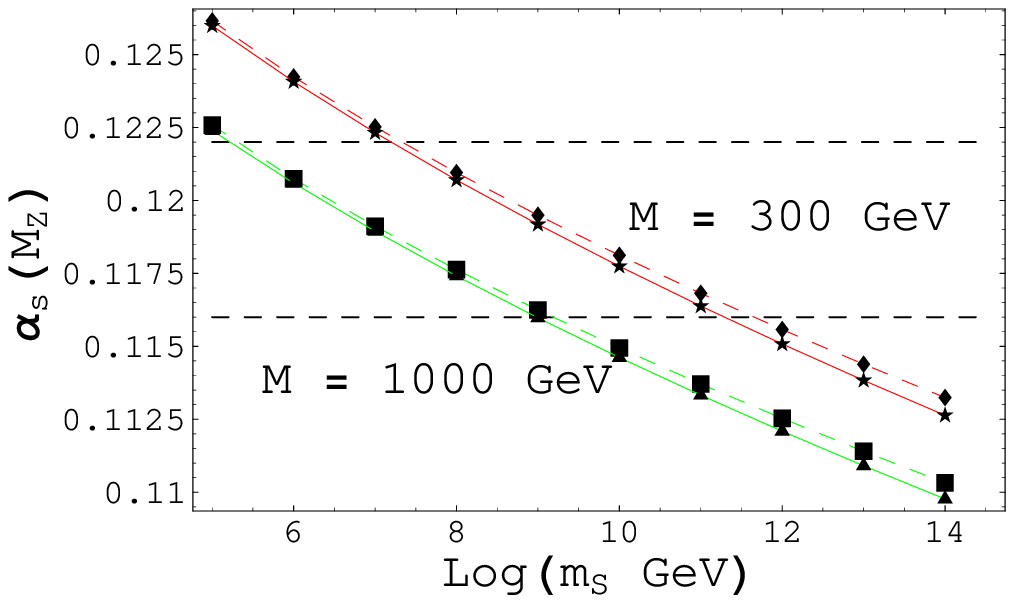}
\caption{The prediction for $\alpha_s(M_Z)$ as a function of the SUSY breaking scale $m_S$. The top graphic is for
the 5D SU(5) model with brane Higgs fields and the bottom graphic is for a 6D SO(10) model with brane Higgs
fields. The horizontal dashed lines show the $1\sigma$ experimental constraint for
$\alpha_s^{exp}(M_Z)$\cite{Eidelman:2004wy}. The solid lines correspond to $\tan(\beta)=50$ and the dashed lines
to $\tan(\beta)=1.5$. The compactification scale is set to $M_c^{\prime}=4\times 10^{14}$ GeV. We assume higgsino
and gaugino mass unification with mass $M$ at the compactification scale. Both $M=300$ GeV and $M=1000$ GeV are
shown.} \label{fig:5Dand6DBraneStrCou}
\end{figure}

\begin{figure}
\includegraphics[width=12cm]{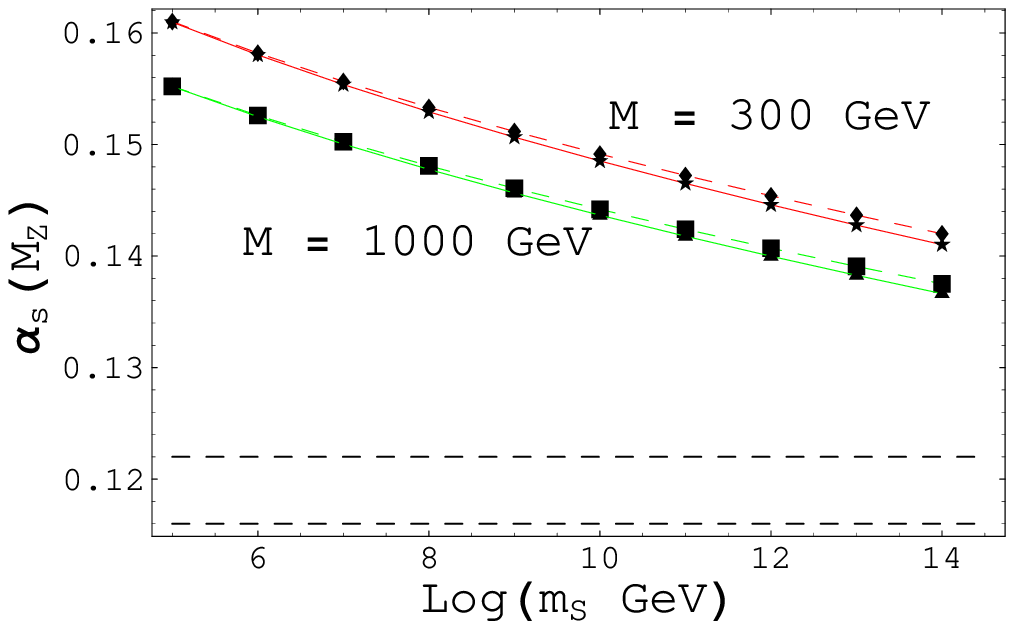}
\includegraphics[width=12cm]{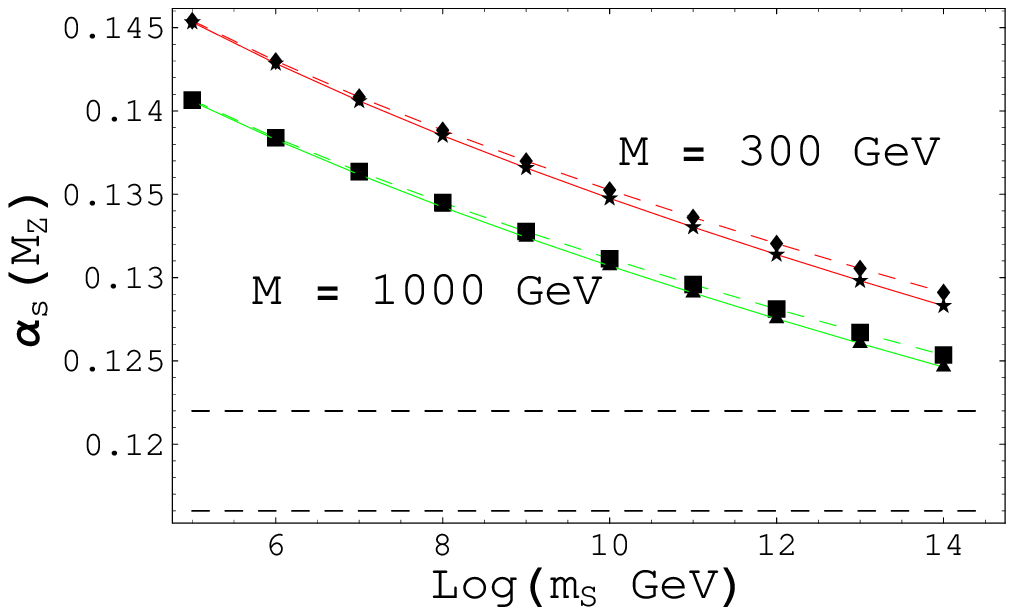}
\caption{The prediction for $\alpha_s(M_Z)$ as a function of the SUSY breaking scale $m_S$. The top graphic is for
the 5D SU(5) model with the Higgs coming from a bulk vector and the bottom graphic is for a 6D SO(10) model with
the Higgs coming from a bulk vector. The horizontal dashed lines show the $1\sigma$ experimental constraint for
$\alpha_s^{exp}(M_Z)$\cite{Eidelman:2004wy}. The solid lines correspond to $\tan(\beta)=50$ and the dashed lines
to $\tan(\beta)=1.5$. The compactification scale is set to $M_c^{\prime}=4\times 10^{14}$ GeV. We assume higgsino
and gaugino mass unification with mass $M$ at the compactification scale. Both $M=300$ GeV and $M=1000$ GeV are
shown.} \label{fig:5Dand6DVectorStrCou}
\end{figure}

Figure \ref{fig:4DStrCou}, shows our two-loop results for unification in the original 4d split-SUSY model of
\cite{Arkani-Hamed:2004fb,Giudice:2004tc,Arkani-Hamed:2004yi}. The behavior of $\alpha_s(M_Z)$ as a function of
$m_S$ and $M$ is clearly well described by eqs. (\ref{AlphaDifferenceA}) and (\ref{AlphaDifferenceB}) and is in
agreement with \cite{Giudice:2004tc}.

Figure \ref{fig:5Dand6DHyperStrCou} shows our results for bulk hypermultiplet Higgs fields in the 5d SU(5) model
of section 2 and a 6d SO(10) model. Our 6d SO(10) model results were obtained rather naively by taking $d=2$ and
$C=8$ in eq. (\ref{StrongCouplingAssumption}) and then adjusting our boundary conditions at $M_c^{\prime }$
accordingly. We expect this change to account for most of the difference in the non-universal running above
$M_c^{\prime}$ and we ignored any additional model dependent details. As expected from eqs.
(\ref{AlphaDifferenceC}) and (\ref{DeltaKKResults}), we can see that the effects of non-universal GUT scale
running and increasing $m_S$ both lower $\alpha_s(M_Z)$ and consequently disfavor split-SUSY with bulk
hypermultiplet Higgs fields. The magnitude of the KK contributions scale like $\frac{1}{d}$, and so the 6d model
is disfavored slightly less. In both cases, $m_S\leq 10^{5}$ GeV and light gauginos and higgsinos are required in
order to match experiment.

Figure \ref{fig:5Dand6DBraneStrCou} displays our unification results for brane Higgs fields in the 5d SU(5) model
and 6d SO(10) model. In these cases, the effect of lowering $\alpha_s(M_Z)$ as $m_S$ is increased is compensated
by the positive contribution to $\alpha_s(M_Z)$ coming from non-universal running above $M_c^{\prime}$. For the 5d
SU(5) model, $m_S$ is favored to be in the range $m_S=10^{10\pm 2}$ GeV, and for 6d SO(10) in the range
$m_S=10^{8\pm 2}$ GeV.

According to eq. (\ref{DeltaKKResults}), the same counter competing effects on $\alpha_s(M_Z)$ occur for the bulk
vector multiplet Higgs case as well. However, the magnitude of the non-universal contribution is four times as
large than in the brane Higgs case and so all but very large $m_S$ is disfavored as is shown in figure
\ref{fig:5Dand6DVectorStrCou}. This problem is less severe for the 6d SO(10) model where the bulk vector multiplet
Higgs case requires at least $m_S=10^{14}$ GeV and heavy gauginos and higgsinos to not be disfavored. So in this
case, the preferred SUSY breaking scale coincides with the compactification scale.

\section{Conclusions}
The split-SUSY scenario offers an interesting new framework for beyond the standard model physics. Not motivated by naturalness as in the MSSM,
unification becomes one of the central motivations for this scenario. Higher dimensional orbifold GUTs offer a particularly compelling
unification framework in which many of the standard problems of SUSY GUTs can be overcome. In this spirit, we incorporated split-SUSY into a 5d
SU(5) orbifold GUT. The primary constraint on these models is a successful prediction of low energy gauge couplings. In generic orbifold GUT
models, non-universal running above the compactification scale alters the low energy unification prediction of $\alpha_s(M_Z)$. The magnitude
and sign of these contributions depends primarily on the bulk geography of the Higgs fields. On the other hand, lifting the scale of
supersymmetry breaking lowers the unification prediction for $\alpha_s(M_Z)$. Our one- and two-loop analysis of gauge coupling unification shows
that split-SUSY favors brane Higgs fields and relatively high scales of SUSY breaking of order $10^{10\pm 2}$ GeV.

\section{Acknowledgements}
P.C.S. would especially like to thank Nima Arkani-Hamed for inspiring this project and for many insightful
discussions and comments. Additional thanks to Natalia Toro for many helpful comments during the completion of
this work. P.C.S. is supported by an NDSEG Fellowship.


\end{document}